\documentclass[prb,showpacs,twocolumn]{revtex4}

\newcommand {\etal}{{\it et al.}}
\newcommand{\COSO}{Cu$_2$OSeO$_3$}

\usepackage{graphicx}
\usepackage{bm}
\usepackage{pifont}
\usepackage{amsmath}

\begin{document}

\title{Formation and rotation of skyrmion crystal in the chiral-lattice insulator Cu$_2$OSeO$_3$}

\author{S. Seki$^1$, J. -H. Kim$^2$, D. S. Inosov$^2$, R. Georgii$^3$, B. Keimer$^2$, S. Ishiwata$^1$, and Y. Tokura$^{1,4,5}$} 
\affiliation{$^1$ Department of Applied Physics and Quantum Phase Electronics Center (QPEC), University of Tokyo, Tokyo 113-8656, Japan \\ $^2$ Max Planck Institute for Solid State Research, Heisenbergstra\ss e 1, 70569 Stuttgart, Germany \\ $^3$ Forschungsneutronenquelle Heinz Maier-Leibnitz (FRM II), Technische Universit\"{a}t M\"{u}nchen, D-85748 Garching, Germany. \\ $^4$  Cross-Correlated Materials Research Group (CMRG) and Correlated Electron Research Group (CERG), RIKEN Advanced Science Institute, Wako 351-0198, Japan \\  $^5$  Multiferroics Project, ERATO, Japan Science and Technology Agency (JST), Tokyo 113-8656, Japan}

\date{}

\begin{abstract}

Small angle neutron scattering experiments were performed on a bulk single crystal of chiral-lattice multiferroic insulator Cu$_2$OSeO$_3$. In the absence of an external magnetic field, helical spin order with magnetic modulation vector $q \parallel \langle 001 \rangle$ was identified. When a magnetic field is applied, a triple-$q$ magnetic structure emerges normal to the field in the A-phase just below the magnetic ordering temperature $T_c$, which suggests the formation of a triangular lattice of skyrmions. Notably, the favorable $q$-direction in the A-phase changes from $q \parallel \langle 110 \rangle$ to $q \parallel \langle 001 \rangle$ upon approaching $T_c$. Near the phase boundary between these two states, the external magnetic field induces a 30$^\circ$-rotation of the skyrmion lattice. This suggests a delicate balance between the magnetic anisotropy and the spin texture near $T_c$, such that even a small perturbation significantly affects the ordering pattern of the skyrmions.

\end{abstract}
\maketitle

\begin{figure}
\begin{center}
\includegraphics*[width=8.5cm]{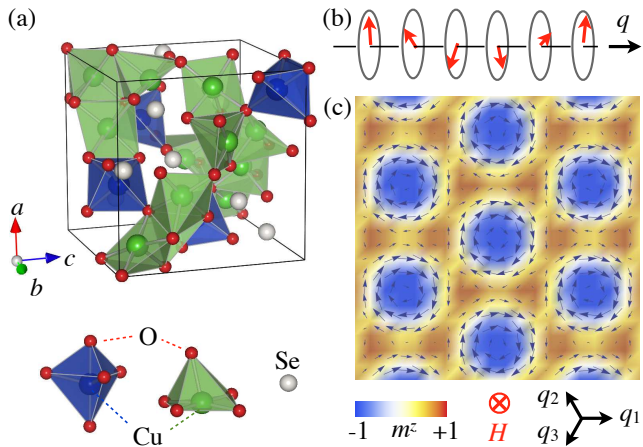}
\caption{(Color Online) (a) Crystal structure of {\COSO} with two distinct Cu$^{2+}$ sites with different oxygen coordination. (b) Proper screw spin order with a single magnetic modulation vector $q$. (c) Skyrmion crystal with triple magnetic modulation vectors $q_1, q_2,$ and $q_3$, which lies within a plane normal to the applied magnetic field ($H$). The background color represents the out-of-plane component of the local magnetic moment ($m_z$).}
\end{center}
\end{figure}

\begin{figure}
\begin{center}
\includegraphics*[width=8.5cm]{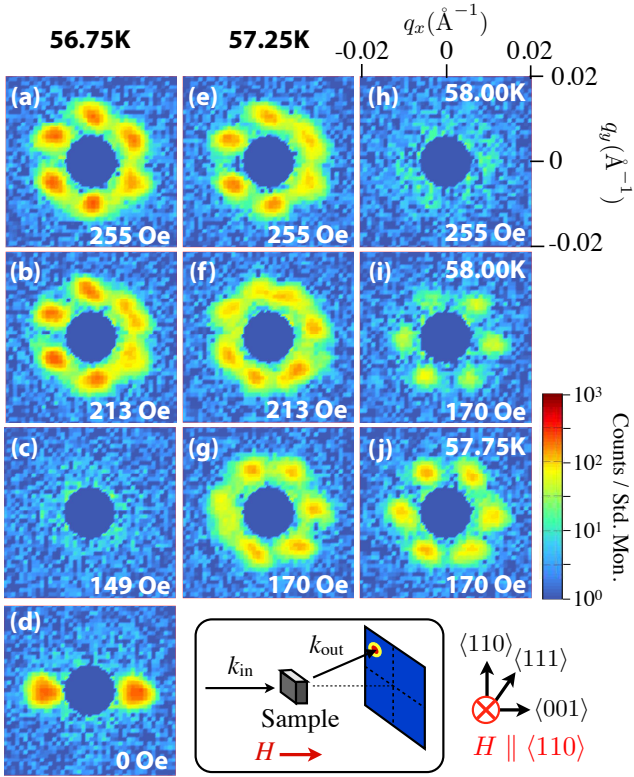}
\caption{(Color Online) Typical SANS profiles for the (110) plane of bulk {\COSO}, measured at various magnetic fields and temperatures. All the data were obtained in the $H$-increasing scans. (a)-(d), (e)-(g), (h)-(i), and (j) represent the ones at 56.75 K, 57.25 K, 58.00 K, and 57.75 K, respectively. The inset shows the measurement setup; $k_\textrm{in}$ and $k_\textrm{out}$ are the incident and scattered neutron wave vectors, respectively. The magnetic field ($H$) is applied along $k_\textrm{in} \parallel \langle 110 \rangle$, i.e. perpendicular to the observed plane.}
\end{center}
\end{figure}

\begin{figure}
\begin{center}
\includegraphics*[width=8.5cm]{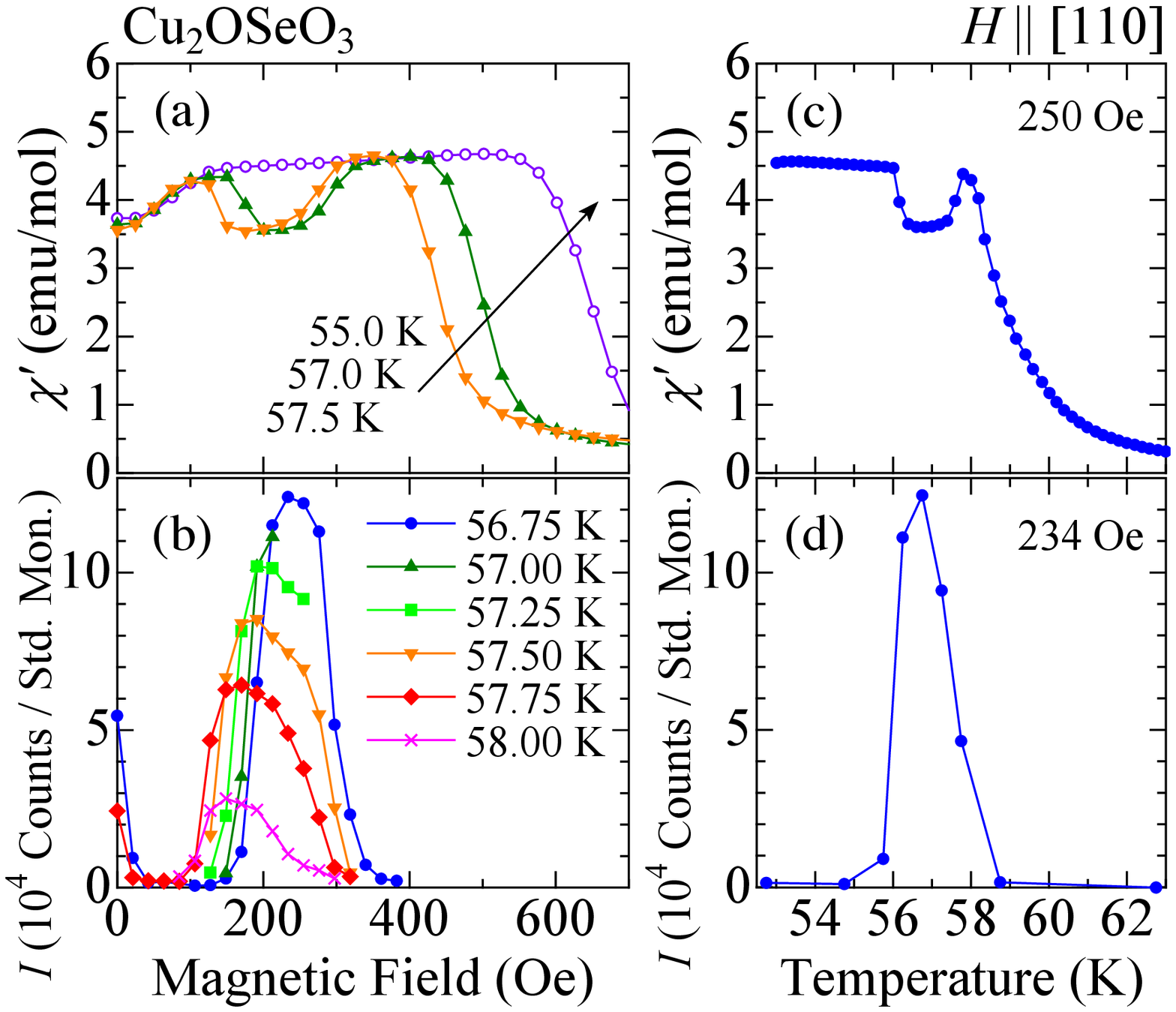}
\caption{(Color Online) Magnetic field ($H \parallel [110]$) dependence of (a) ac magnetic susceptibility $\chi '$ and (b) integrated intensity $I$ of magnetic reflections obtained from SANS profiles. Temperature dependence of $\chi '$ and $I$ is also plotted in (c) and (d).}
\end{center}
\end{figure}

\begin{figure}
\begin{center}
\includegraphics*[width=8.5cm]{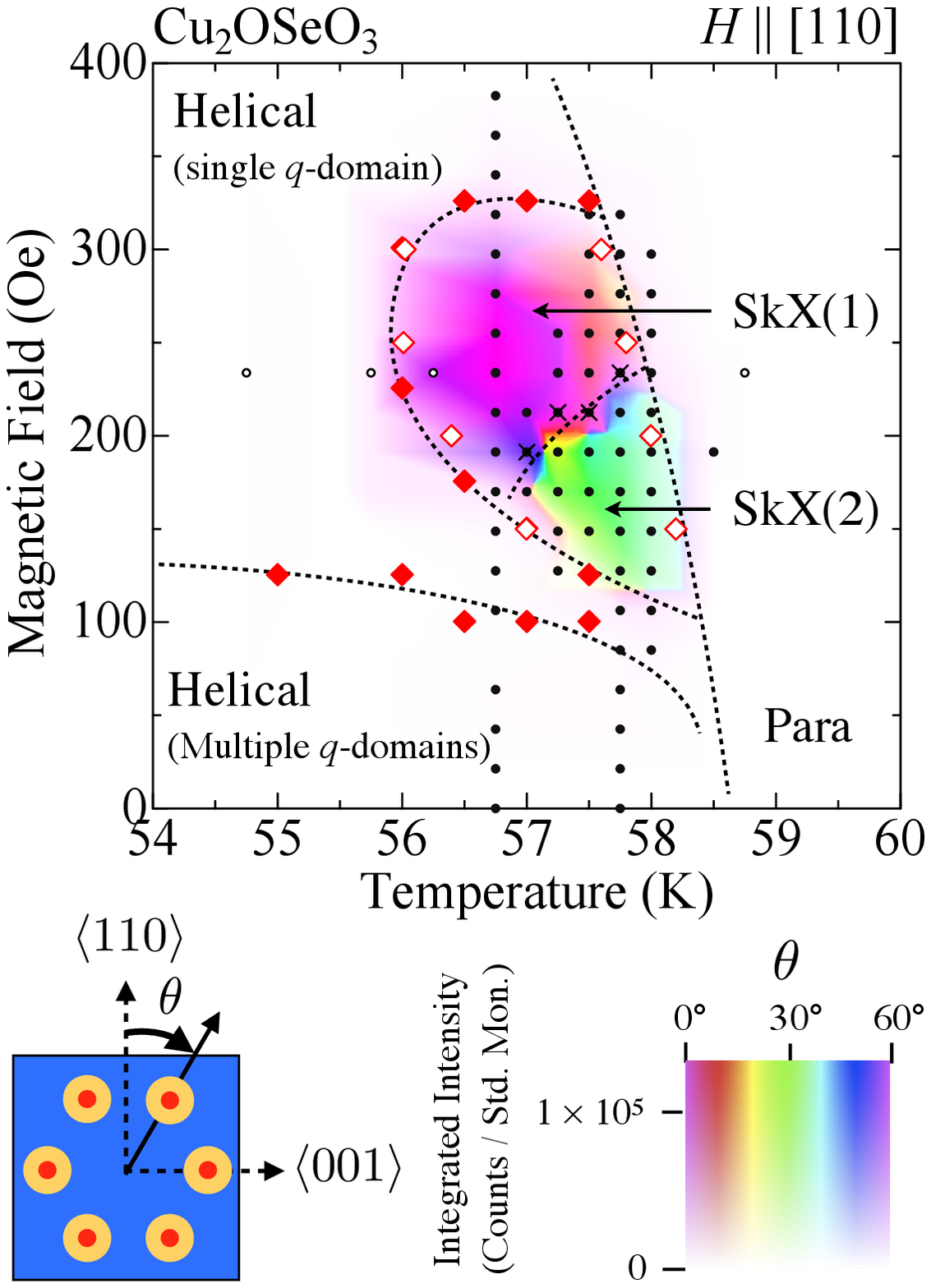}
\caption{(Color Online) Temperature ($T$) and magnetic field ($H$) phase diagram obtained for the bulk {\COSO} with $H \parallel [110]$. Diamonds represent the anomalies in $\chi '$-profile, and circles indicate the data points where SANS measurements were performed. In both cases, closed and open symbols are obtained from the $H$- and $T$-scans, respectively. The background color indicates the properties of SANS profile at each data point; color saturation represents the integrated intensity of six-fold magnetic reflections, and the hue stands for the rotation angle ($\theta$) of the magnetic Bragg position with respect to the $\langle 110 \rangle$ direction (see the bottom panel). Typical SANS profiles at selected data points are shown in Figs. 2 (a)-(j). In the data points with cross symbols, the mixed SANS pattern as shown in Fig. 2 (f) was observed. Here, the A-phase consists of SkX(1) and SkX(2); The former corresponds to the skyrmion crystal state favoring $q \parallel \langle 110 \rangle$, and the latter does the one favoring $q \parallel \langle 001 \rangle$.}
\end{center}
\end{figure}

The skyrmion is a topologically stable particle-like object, which appears as a vortex-like swirling spin texture in chiral-lattice magnets.\cite{Skyrme, SkXTheory, SkXBogdanov} In metallic compounds, conduction electrons flowing over skyrmion spin texture feel a fictitious magnetic flux arising from the spin chirality of the skyrmion or skyrmion number, thereby producing the topological Hall effect.\cite{THETheory, THEMnSi} On the other hand, the electric current can drag the skyrmion via a spin transfer torque, while the motion of the skyrmion generates a transverse electromotive force as the emergent electric field.\cite{STTMnSi, EmergentMnSi} Such electric controllability, as well as its particle-like nature with nanometric size, highlights skyrmion as a unique magnetic object with potential applications in spintronics and high-density magnetic storage devices. In chiral-lattice magnets, the magnetic interaction acting on two neighboring spins ($\vec{S}_i$ and $\vec{S}_j$) is mainly composed of two terms; the $\vec{S}_i \cdot \vec{S}_j$-like (ferromagnetic) exchange interaction and the $\vec{S}_i \times \vec{S}_j$-like Dzyaloshinskii-Moriya interaction. The latter term tends to orient the neighboring spins in a canted fashion, and hence the helical spin structure with single modulation vector $q$ is generally favored (Fig. 1(b)). In a limited temperature ($T$) and magnetic field ($H$) window, however, skyrmion spin state can dominate over the helical ground state. In most cases, skyrmions crystallize to form a triangular lattice within the plane normal to the magnetic field (Fig. 1(c)), while they can also exist as independent particles.\cite{NeutronMnSi, TEMFeCoSi}

Magnetic skyrmions have been first experimentally discovered in intermetallic compounds such as MnSi,\cite{NeutronMnSi} Fe$_{1-x}$Co$_x$Si,\cite{TEMFeCoSi, NeutronFeCoSi} and FeGe.\cite{TEMFeGe} These compounds share the same B20 structure with chiral cubic symmetry, and commonly show the helimagnetic gound state. In the bulk form, their magnetic phase diagram includes an $H$-induced narrow region, called "A-phase", just below the magnetic ordering temperature $T_c$.\cite{MnSi_APhase} Small angle neutron scattering (SANS) studies in the A-phase revealed that the three magnetic modulation vectors form an angle of $120^\circ$ with respect to each other within a plane normal to the applied $H$, which suggests the formation of a triangular lattice of skyrmions (Fig. 1(c)).\cite{NeutronMnSi, NeutronFeCoSi} For thin films, such a skyrmion crystal (SkX) has indeed been observed in real-space through Lorentz transmission electron microscopy (TEM), while the SkX in the thin-film form is stabilized over much wider $T$- and $H$-range than the bulk case.\cite{TEMFeCoSi, TEMFeGe}

Recently, through Lorentz TEM imaging of a thin film, a SkX has also been observed for the insulating compound {\COSO}.\cite{COSO_SkX} Its crystal structure belongs to the same chiral cubic space group $P2_13$ as the B20 compounds, but with a very different atomic arrangement (Fig. 1(a)).\cite{COSO_First, COSO_Structure} The magnetic phase diagram of {\COSO}, like that of the B20 compounds, is radically different between bulk and thin film. For bulk specimens, previous powder neutron scattering and NMR studies suggested a three-up one-down type ferrimagnetic arrangement between two inequivalent Cu$^{2+}$ ($S=1/2$) sites,\cite{COSO_Dielectric, COSO_NMR, COSO_Ferri} while ac magnetic susceptibility measurements indicated a more complex magnetic phase diagram characterized by an A-phase similar to that of the B20 compounds.\cite{COSO_SkX} The magnetic structure in the A-phase of bulk {\COSO} has not yet been resolved, although the analogy with the B20 compounds implies the possible formation of a SkX in this region. Notably, the bulk form of {\COSO} exhibits a magnetically induced electric polarization in the A-phase,\cite{COSO_SkX, COSO_Dielectric} and thus is considered as a multiferroic (i.e. material with both magnetic and dielectric orders).\cite{ME_Review_Fiebig, ME_Review_Cheong, ME_Review_Seki} Such magnetically-induced ferroelectrics often allow electric field control of magnetism,\cite{ME_PN_LiCu2O2, ME_QVec, ME_MagneticDomain} and the observed magnetoelectric coupling in the A-phase implies the possible manipulation of skyrmions by external electric fields in insulators. To establish the coupling between electric polarization and skyrmions, a detailed investigation of spin textures in bulk {\COSO} is urgently required.

In this paper, we study the long-wavelength magnetic structure of bulk {\COSO } by SANS measurements. We identified the helical nature of the magnetic ground state, and further discovered the formation of a skyrmion lattice in the bulk A-phase. This establishes the coupling between skyrmion spin texture and electric polarization, which promises the potential manipulation of skyrmions by external electric fields, not electric currents. Interestingly, the skyrmion lattice shows a $30^\circ$-rotation within the plane of observation as a function of temperature and magnetic field. This finding suggests an extreme sensitivity of the skyrmion ordering pattern to perturbation.

Single crystals of {\COSO} were grown by the chemical vapor transport method.\cite{COSO_InfraRed} AC magnetic susceptibility $\chi '$ (at 700 Hz) was measured with a SQUID magnetometer. SANS measurements were performed at the diffractometer MIRA at FRM II using neutrons with a wavelength $\lambda \sim 5.1$ \AA. Backgrounds were determined at high temperatures and subtracted accordingly. A magnetic field was applied parallel to  incident neutron beam using Helmholtz coils (for setup see Ref. 8). Data were recorded for fixed sample orientations, following rocking scans with respect to a vertical axis of typically $\pm 1.8 ^\circ$. All measurements were performed on the same as-grown crystal with the size of 8 mm $\times$ 5 mm $\times$ 3 mm, and the magnetic field was always applied normal to the (110) widest surface. All $H$-scans were performed in the $H$-increasing process after zero field cooling (ZFC), and the $T$-scans were performed in the warming process after ZFC and subsequent application of $H$.

First, we focus on the behavior at 56.75 K, which is close to $T_c \sim 58$ K. Figure 2(d) indicates the SANS profile in zero magnetic field. We found two magnetic Bragg reflections with $q \parallel \langle 001 \rangle$, which suggests that the magnetic ground state of bulk {\COSO} is not simply ferrimagnetic, but rather helically modulated. The observed wave number $q \sim 0.010$ \AA$^{-1}$ corresponds to the modulation period of $\sim 63$ nm, which roughly agrees with the helical modulation period $\sim 50$ nm previously reported for the thin film.\cite{COSO_SkX, CommentPeriod} This value is much longer than the Cu-Cu atomic distance and thus compatible with a local ferrimagnetic arrangement. The direction and wave number of $q$ remain unchanged down to 10 K (not shown). Due to the high symmetry of the cubic lattice, the equal population of $q \parallel  \langle 001 \rangle$ domains will coexist. Notably, the propagation vector of the helimagnetic ground state is different between bulk ($q \parallel \langle 001 \rangle$) and thin film ($q \parallel \langle 110 \rangle$)\cite{COSO_SkX} for {\COSO}. Recent theories suggest that the stable $q$-direction in the helical ground state depends on the dimensionality of the system;\cite{SkXGround} They predict $q \parallel \langle 110 \rangle$ for a two-dimensional (2D) system, while $q \parallel \langle 001 \rangle$ or $q \parallel \langle 111 \rangle$ for a 3D system. The observed $q$ directions in the bulk (3D) and thin film (2D) forms of {\COSO} agree well with these predictions.

In the following, we discuss the development of SANS profiles under magnetic field applied along the out-of-plane $\langle 110 \rangle$ direction. Previous Lorentz TEM measurements on a thin film suggested a screw-type helical ground state, where the spins rotate within a plane perpendicular to $q$ (Fig. 1(b)). Since antiferromagnetically aligned spins tend to orient normal to $H$, the applied magnetic field should favor the $H \parallel q$ relationship and reorient $q$ along the out-of-plane direction in the present SANS setup. Figure 3 (b) indicates the $H$-dependence of the integrated intensity of the magnetic reflections obtained from the SANS profiles. Above 50 Oe, the magnetic reflection vanishes from the plane of observation as shown in Fig. 2(c), which is consistent with such an $H$-induced reorientation of $q$ in the helical state.

Upon increasing the magnetic field further, magnetic reflections in the SANS profile emerge again at 200 Oe $< H <$ 300 Oe (Fig. 3(b)). The corresponding $H$-dependence of the ac magnetic susceptibility $\chi '$ shows a clear dip anomaly in this field region (Fig. 3 (a)), which signals the transition into the A-phase.\cite{COSO_SkX, B20Many} Figures 2 (a) and (b) indicate the SANS profiles at 255 Oe and 213 Oe, respectively. We observed a six-fold symmetric pattern of magnetic Bragg spots within a plane normal to the applied $H$, which evidences the formation of a skyrmion crystal in the A-phase of bulk {\COSO} (Fig. 1 (c)).\cite{NeutronMnSi, NeutronFeCoSi} Three magnetic modulation vectors form an angle of 120$^\circ$ with respect to each other, and one of them satisfies the $q \parallel \langle 110 \rangle$ relationship. Interestingly, the favorable $q$ direction in the SkX state of bulk {\COSO} depends on temperature. Figure 2 (j) shows the SANS profile at $T = 57.75$ K with $H = 170$ Oe. The six-fold magnetic reflections verify the skyrmion lattice formation, while the SANS pattern rotates by $30^\circ$ with respect to the one at 56.75 K. In this case, one of the magnetic modulation vectors rather favors the $q \parallel \langle 001 \rangle$ relationship.

Based on the various $T$- and $H$-scans of $\chi '$ and the SANS profiles, we have summarized the $H$-$T$ phase diagram for bulk {\COSO} with $H \parallel \langle 110 \rangle$ (Fig. 4). The anomalies in the $\chi '$ profile are indicated with diamonds, and the A-phase is identified in the narrow region of 56 K $< T <$ 58 K and 100 Oe $< H <$ 350 Oe. The background color in the phase diagram indicates the nature of the SANS profiles at each data point. The color saturation represents the integrated intensity of the six-fold magnetic reflections, and the hue reflects the rotation angle of the magnetic modulation vector with respect to the $\langle 110 \rangle$ direction (see the bottom panel of Fig. 4). The emergence of six-fold magnetic reflections always coincides with the dip anomaly in the $\chi '$-profiles (Figs. 3 (a)-(d)), which identifies the A-phase as the SkX state in the bulk crystal. Our SANS measurements have further revealed that the A-phase is divided into two regions; the SkX(1) phase favoring $q \parallel \langle 110 \rangle$, and the SkX(2) phase favoring $q \parallel \langle 001 \rangle$. Near the phase boundary, we can even switch these two SkX states by the external magnetic field. Figures 2 (e)-(g) indicate the $H$-dependence of the SANS profiles at $T=57.25$ K. The transition from SkX(2) to SkX(1) is induced as $H$ increases from 170 Oe to 255 Oe, and a mixed pattern of these two profiles due to phase coexistence is observed at the phase boundary (213 Oe). Note that neither $\chi '$ nor the integrated intensity of the magnetic reflections show any noticeable anomaly at the transition between these two SkX states (Figs. 3(a) and (b)). The magnetic modulation period remains unchanged among the helical and the two SkX spin states.

For metallic MnSi, a rotation of the skyrmion lattice by up to 10$^\circ$ has recently been reported under the combination of a temperature gradient $\Delta T$ and an electric current $j$.\cite{STTMnSi} Its rotation angle becomes larger as the magnitude of $\Delta T$ and $j$ increases, and the origin of this phenomenon is ascribed to a Berry phase collection by conduction electrons flowing over skyrmions. This scheme is clearly inapplicable to the present case, since {\COSO} is an insulator and the rotation angle seems to be fixed to 30$^\circ$ without any current. According to a recent phenomenological Ginzburg-Landau analysis, the pinning of the $q$-direction in the SkX state is governed by a magnetic anisotropy of sixth order in spin-orbit coupling.\cite{NeutronMnSi, NeutronFeCoSi} For example, the terms like $\sum_q (q^6_x + q^6_y + q^6_z) |\vec{m}_q|^2$ and $\sum_q (q^4_x q^2_y + q^4_y q^2_z + q^4_z q^2_x) |\vec{m}_q|^2$ favor $q \parallel \langle 110 \rangle$ or $q \parallel \langle 001 \rangle$ depending on the sign of their prefactor. Here, $\vec{m}_q$ is the Fourier component of the spatially dependent local magnetization $\vec{M}(\vec{r})$. The observed rotation of SkX suggests that the pinning of $q$ is very weak, and that the SkX(1) and SkX(2) states are nearly degenerate. Their relative stability is reversed near $T_c$, perhaps because the enhancement of thermal fluctuation\cite{NeutronMnSi} and/or higher harmonics in spin texture\cite{NeutronMnSi, MnSiHigher} near $T_c$ provide nonmonotonic contributions to the anisotropy terms. While a further theoretical analysis is necessary to fully clarify its origin, the current observation of a SkX rotation indicates an extreme sensitivity of the skyrmion ordering pattern against external field or perturbation.

Combined with the previous reports for the thin film,\cite{COSO_SkX} the current discovery of long-wavelength spin textures in a bulk crystal fully establishes {\COSO} as an insulating counterpart of the B20 compounds. Both systems have a chiral cubic lattice, and their magnetic phase diagrams are characterized by the helical ground state as well as the $H$-induced SkX state. The stability of the SkX state depends on the sample dimension and it shrinks into the narrow A-phase in the bulk limit.\cite{TEMFeGe, COSO_SkX} These similarities are rather surprising considering the significant differences between the two systems, such as the electronic structure (metal or insulator), the local spin arrangement (ferrimagnetic or ferromagnetic), and the atomic arrangement in the crystal. This finding strongly suggests that chiral cubic ferro/ferrimagnets may ubiquitously host skyrmions, irrespective of other details. In addition, the identification of skyrmions in the A-phase of bulk {\COSO} establishes the coupling between the skyrmion spin texture and the previously reported electric polarization,\cite{COSO_SkX} which promises the manipulation of skyrmions by external electric fields. Since an electric field in insulators causes negligible joule heat loss compared to the current-driven approach in metals, the currently established {\it magnetoelectric} skyrmion may be useful for the design of unique spintronic devices with high energy efficiency.

In summary, we have performed a detailed small-angle neutron scattering study of the bulk form of the chiral-lattice multiferroic insulator {\COSO}. The helical nature of the magnetic ground state, as well as the formation of a skyrmion lattice in the A-phase, have been successfully identified. This establishes a coupling between electric polarization and skyrmions in the A-phase, which implies the possibility of manipulating skyrmions by external electric fields. Notably, the favorable $q$-direction in the skyrmion state changes as $T_c$ is approached. Near the phase boundary, the external magnetic field induces a $30^\circ$-rotation of the skyrmion crystal. This finding demonstrates the delicate balance between the magnetic anisotropy and spin texture near $T_c$, and suggests that the skyrmion ordering pattern may be controlled by various external stimuli.

The authors thank T. Arima, N. Nagaosa, M. Mochizuki, N. Kanazawa, X. Z. Yu, K. Shibata, M. Rikiso, and all people at FRM II for enlightening discussions and experimental helps. This work was partly supported by Grants-In-Aid for Scientific Research (Grant No. 20340086, 2010458) from the MEXT of Japan, and FIRST Program by the Japan Society for the Promotion of Science (JSPS) and by the DFG under SFB/TRR 80.


\begin{thebibliography}{100}
\bibitem{CommentPeriod} A slightly shorter helical modulation period or skyrmion lattice constant as observed in the Lorentz TEM film sample ($\sim 100$ nm-thick)\cite{COSO_SkX} may reflect the contribution from the magnetic dipole-dipole interaction that becomes important in the thin film form of specimen; the dipolar interaction is anticipated to stabilize the skyrmion formation and change the modulation period as well.\cite{BubbleOne, BubbleTwo}
\end{thebibliography}
\end{document}